\documentclass[preprint]{rmaa}

\title{A VLA Archive Observation of the 
Youngest Known Galactic Supernova Remnant G1.9+0.3}

 \author{
 Yolanda G\'omez,\altaffilmark{1}
         and Luis F. Rodr\'\i guez,\altaffilmark{1}} 
  \altaffiltext{1}{Centro de Radioastronom\'{\i}a y Astrof\'{\i}sica, Universidad
Nacional Aut\'onoma de M\'exico, Campus Morelia} 

\fulladdresses{
\item Yolanda G\'omez and Luis F. Rodr\'\i guez: Centro de Radiostronom\'{\i}a
y Astrof\'\i sica, 
UNAM, A. P. 3-72, 
(Xangari), 58089 Morelia, Michoac\'an, M\'exico
(y.gomez; l.rodriguez@astrosmo.unam.mx).
}

\shortauthor{G\'omez \& Rodr\'\i guez}
\shorttitle{A VLA Archive Observation of G1.9+0.3}

\SetVolume{50} \SetFirstPage{001} \SetYear{2009}
\ReceivedDate{\today} 
\AcceptedDate{Year Month Day} 

\resumen{Presentamos el an\'alisis de una observacion de archivo
sin publicar
hecha con el Very Large Array a 1.49 GHz en 1989 hacia la remanente
de supernova G1.9+0.3, la m\'as joven conocida en la Galaxia. 
Esta observaci\'on concuerda con la evoluci\'on temporal
tama\~no angular reportada anteriormente.
Derivamos una tasa de expansi\'on angular de $0.46 \pm 0.11$\%
por a\~no y estimamos una edad de $220 \pm_{45}^{70}$
a\~nos comparando las im\'agenes de 1985 y 1989.
}
 
\abstract{We present the analysis of an unpublished VLA archive observation made
at 1.49 GHz in 1989 toward the supernova remnant G1.9+0.3, the youngest
such Galactic object known. 
This observation agrees with the time evolution
in angular size
previously reported.
We derive an expansion rate of $0.46 \pm 0.11$\% per year and an age of 
$220 \pm_{45}^{70}$ yr for the remnant by comparing the 1985 and 1989 images.
}
      

\begin{document}

\maketitle

\section{Introduction}

In a remarkable result, Green et al. (2008) and Reynolds et al. (2008) 
reported the fast expansion (13,000$\pm$1,000 km s$^{-1}$ at an
assumed distance of 8.5 kpc) of the compact supernova remnant G1.9+0.3,
concluding that, with an age of order 100 years, it is the youngest such
object in the Galaxy. Obviously, any past observation of this source 
becomes now particularly valuable since they will help to characterize
better the parameters of supernova remnants in a very young stage. 

In this paper we present the analysis of archive 1.49 GHz 
continuum observations made with the Very
Large Array (VLA) in the B configuration toward G1.9+0.3 
in two epochs: 1985 April 16 (epoch 1985.29) and 1989 April 29
(epoch 1989.33). The first epoch has
been analyzed and discussed in detail by Green et al. (2008) and Reynolds et al. (2008)
since it constitutes the first epoch of their expansion
measurements. Reynolds et al. (2008) compared the VLA 1.49 GHz observations of
1985 April 16 with Chandra observations taken during 2007 February 10 and March 3
with the ACIS-S CCD camera, while Green et al. (2008) compared the same 1985
VLA observations with new VLA observations of similar angular resolution $\sim 10{''} \times
4{''}$, but taken at a different frequency (4.86 GHz) on 2008 March 12.  

Additional radio data for G1.9+0.3 has been presented by De Horta et al. (2008),
who discuss observations made in 1993 using the Australia Telescope Compact Array (ATCA)
at 6 cm and confirm the expansion rate of $\sim$0.65\% per year
found by Green et al. (2008) between 1985 and 2008.
However, the De Horta et al. (2008)
results are derived from images with quite different uv coverage.
Murphy et al. (2008) present a radio light curve for G1.9+0.3 based on 25 epochs of 
843 MHz observation with the Molonglo Observatory Synthesis Telescope, spanning 20 yr from 1988 
to 2007. They find that the flux density has increased at a rate of 
$1.22\pm_{0.16}^{0.24}$\% per year, supporting the suggestion of
Green et al. (2008) that G1.9+0.3 is 
undergoing a period of magnetic field amplification.

In this paper we present the analysis of VLA archive
data taken in 1989 April 29 with the same
frequency and angular resolution than the data of 1985 April 16. To our knowledge,
the 1989 data is unpublished. Our analysis 
allows a more reliable comparison since the data has the same
frequency and angular resolution, but it has the disadvantage that the time baseline is
small, only 4.04 years. 
Even when the 1989 data is constituted of only two 3-minute integrations separated
by 1.5 hours it has excellent quality and allows a comparison with the previous
epoch. 

\section{Data Reduction}

The archive data
were edited and calibrated using the software package Astronomical Image
Processing System (AIPS) of NRAO. Cleaned maps were obtained using the
task IMAGR of AIPS and the ROBUST parameter
(Briggs 1995) of this task set to 5, to optimize sensitivity.
The observational parameters of the two epochs are given in Table 1.

\begin{table*}[htbp]
\small
  \setlength{\tabnotewidth}{2.0\columnwidth} 
  \tablecols{6} 
  \caption{1.49 GHz Archive Data for G1.9+0.3}
  \begin{center}
    \begin{tabular}{lccccc}\hline\hline
                      &         & Time$^a$ & Phase & Bootstrapped & Beam  \\
Epoch                 & Project & (min) & Calibrator & Flux(Jy) & Angular Size$^b$ \\ 
\hline
1985 April 16 (1985.29) & AG184   & 25 & B1829-106 & 0.92$\pm$0.01 & $10\rlap.{''}6 \times 5\rlap.{''}6;~\ -6^\circ$  \\
1989 April 29 (1989.33) & AB515   & 6 & B1748-253 & 1.21$\pm$0.01 & $10\rlap.{''}6 \times 5\rlap.{''}1;~  -14^\circ$  \\
\hline\hline
\tabnotetext{a}{Integration time on source.}
\tabnotetext{b}{Major axis $\times$ minor axis; position angle for images with ROBUST = 5.
The final images were made with a restoring beam of $12\rlap.{''}0 \times 6\rlap.{''}5;~PA ~=~0^\circ$.}
    \label{tab1}
    \end{tabular}
  \end{center}
\end{table*}

\section{Results}

\subsection{Flux Densities}

The flux densities obtained by us at 1.49 GHz are 0.565$\pm$0.078 Jy and 0.646$\pm$0.046 Jy
for the epochs 1985.29 and 1989.33,
respectively.
These flux densities were obtained from the images using the task IMSTAT of
AIPS and estimating the errors following
Beltr\'an et al. (2001).
However, these data were taken in the B configuration at 1.49 GHz, where the
largest angular scales detectable by the array
are only a factor of 2 larger than G1.9+0.3.
We then expect to have poor sampling of the uv plane close
to the zero spacing and that the flux densities reported
above are underestimating the real values.
Additional measurements at 1.4 GHz (0.748$\pm$0.038 Jy; epoch 1996.47; Condon et al. 1998)
and at 1.425 GHz (0.935$\pm$0.047 Jy; epoch 2008.20; Green et al. 2008)
are reported in the literature. 

\subsection{Images}

\begin{figure*}
\centering
\includegraphics[scale=0.32, angle=0]{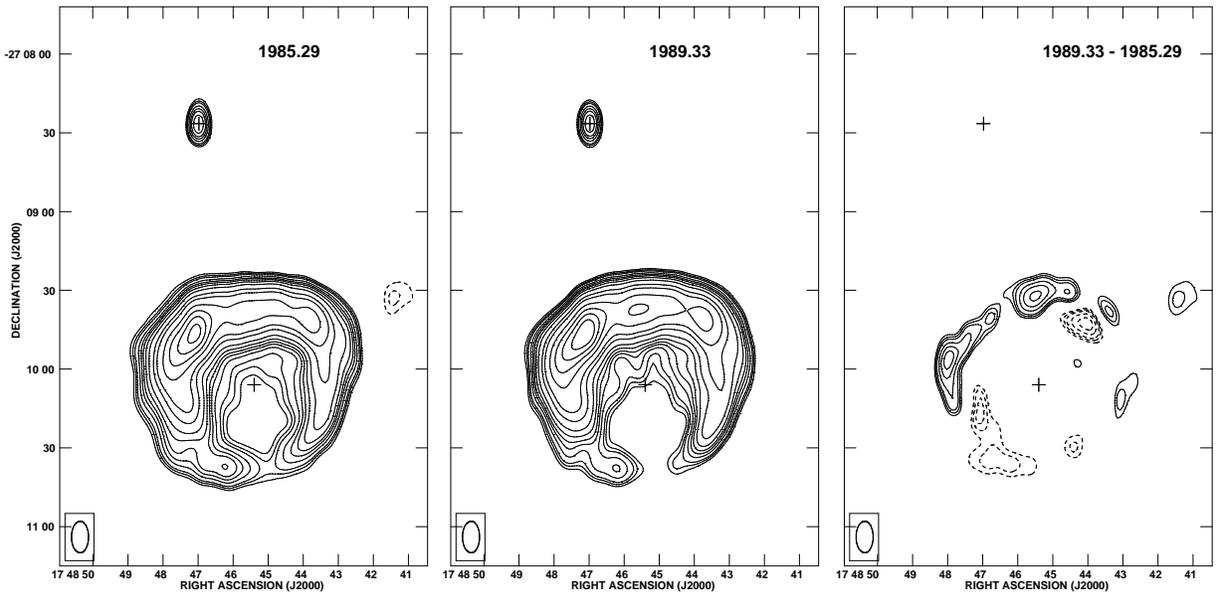}
 \caption{Contour images of the 1.49 GHz continuum
emission from G1.9+0.3 for 1985.29 (left) and 1989.33 (right).
The difference image, taken to be the 1989.33 image minus the
1985.29 times 1.05, is shown in the right panel.
The contours are 10, -8, -6,-5,-4, 4, 5, 6, 8, 10, 12, 15, 20, 30,
40, 60, 80, 100, and 120
times 0.4 $\mu$Jy beam$^{-1}$.
The cross in the northern part of the images marks the position of a field source.
The cross in the southern part of the images marks the centroid
of the radio emission associated with G1.9+0.3, from Green et al. (2008).
The half power contour of the restoring beam 
($12\rlap.{''}0 \times 6\rlap.{''}5;~PA ~=~0^\circ$),
is shown in the bottom left corner of the panels.}
  \label{fig1}
\end{figure*}

To compare the angular size of the SNR 
at the two epochs we made images with the same restoring beam of
$12\rlap.{''}0 \times 6\rlap.{''}5;~PA ~=~0^\circ$. This value is comparable to
the synthesized beams of the two epochs (see Table 1). 
We show in Figure 1 the images of G1.9+0.3 for the two epochs (1985.29 and 1989.33), 
as well as a difference
image (1989.33 - 1985.29). The difference image shows evidence of expansion,
with positive contours in the outer parts of the remnant and
negative contours in the inner parts.  
Unfortunately, the lack of azimutal symmetry 
of the nebula does not produce a clear shell in the difference image,
as observed for example in some planetary nebulae (e. g. Guzm\'an et al. 2006). 
Following Green et al. (2008) we averaged the emission 
over all azimuths, using the task IRING of AIPS.
The central position of these rings was the centroid
of the radio emission associated with G1.9+0.3,
$\alpha(2000) = 17^h~48^m~45\rlap.{^s}4; \delta(2000) = -27^\circ~10'~06{''}$,
from Green et al. (2008).
The shell profiles for
the two epochs, normalized to the peak value of 1989.33
by multiplying the 1985.29 profile by 1.05, are shown in Fig. 2.
The 1989.33 profile is slightly displaced to larger radii
with respect to the 1985.29 profile.
In Fig. 2 we also show the difference of the two profiles,
that shows the S-shaped profile indicative of expansion.
Analyzing these profiles under the assumption that they are Gaussian,
we find a displacement of $0\rlap.{''}57 \pm 0\rlap.{''}14$ between the two epochs. 
Since the peak of the shell is expected to have a radius of $31{''}$ in 1989
(Green et al. 2008), this implies an expansion rate
of $0.46 \pm 0.11$\% per year and an age of $220 \pm_{45}^{70}$ yr for the remnant
(assuming a constant expansion velocity and with respect to the 1989 epoch).
This age is somewhat larger than the value of 150 yr estimated
by Green et al. (2008). Additional observations are required to better estimate
the age of this SNR, but it is certainly a very young object.

This is the first determination of the expansion of the SNR made from data at
the same frequency, since the determination of Reynolds et al. (2008) was made
comparing VLA (radio) and Chandra (X-ray) images and that of Green et al. (2008) was
made comparing a 1.49 GHz image with a 4.86 GHz image (both made with the VLA). 

The NW extension found by Green et al. (2008) in their 2008 image made at
4.9 GHz is not present in the 1989 image made at 1.49 GHz and presented here
(Fig. 1). This implies that the extension must have appeared after 1989.

\begin{figure}
\centering
\includegraphics[scale=0.4, angle=0]{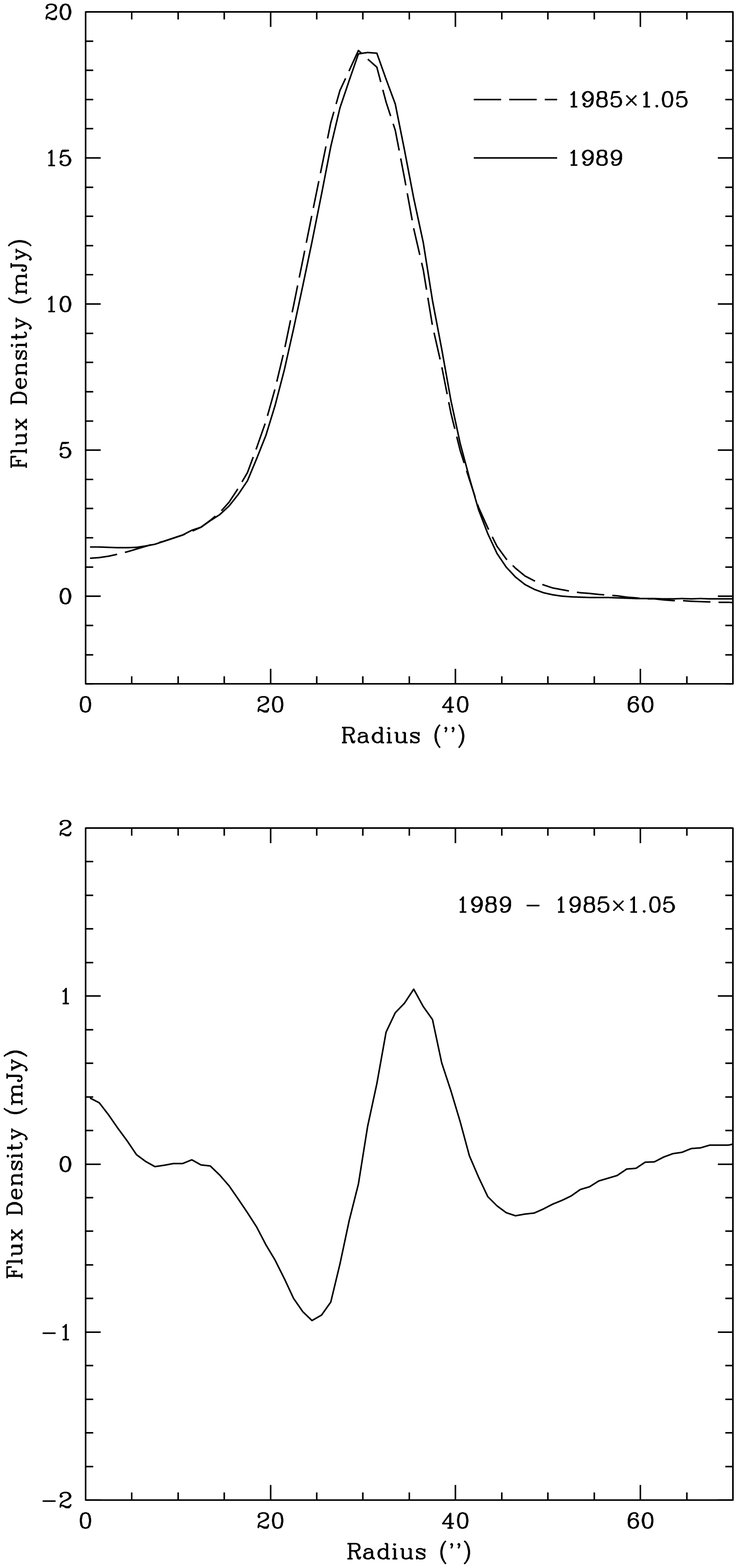}
 \caption{(Top) Shell profiles for the 1.49 GHz continuum
emission from G1.9+0.3 for 1985.29 (dashed line) and 1989.33 (solid
line). The peak intensities have been normalized
by multiplying the 1985.29 profile by 1.05.
Bottom: Difference of the shell profiles.
}
  \label{fig2}
\end{figure}

\section{Conclusions}

Our main conclusions are as follows.

1) We present the analysis of VLA archive data of the young supernova G1.9+0.3
taken at 1.49 GHz in 1989. 

2) This data point agrees well with the evolution
observed in observations taken at
the same or similar frequencies at other
epochs and confirms the angular
expansion previously reported.

3) We derive an expansion rate of $0.46 \pm 0.11$\% per year and an age of 
$220 \pm_{45}^{70}$ yr for the remnant from the 1985 and 1989 data.
This is the first determination of the expansion of the SNR made from data at
the same frequency, since the determination of Reynolds et al. (2008) was made
comparing VLA (radio) and Chandra (X-ray) images and that of Green et al. (2008) was
made comparing a 1.49 GHz image with a 4.86 GHz image (both made with the VLA). 


\acknowledgments
We thank an anonymous referee for valuable comments that
led to an improved version of this paper.
We acknowledge Jane Arthur for calling our attention to this
object and for helpful comments. We are thankful for the support
of DGAPA, UNAM, and of CONACyT (M\'exico).
This research has made use of the SIMBAD database, 
operated at CDS, Strasbourg, France.


\end{document}